# Social Capital and Persistence in Computer Science of Google's Computer Science Summer Institute (CSSI) Students

Marjan Naghshbandi[1], Sharon Ferguson[1], Alison Olechowski[1]
[1]University of Toronto

## Abstract

While a lucrative and growing field, low levels of gender and racial diversity in Computer Science (CS) remain prevalent. Education and workforce support programs with the intention to promote underrepresented students' persistence in CS exist, which teach CS skills, inform of career options, and grow students' network in CS. Studies have demonstrated these programs' effectiveness as it relates to changes in affective outcomes, such as participants' confidence in CS skills and attitudes towards CS jobs. However, the longitudinal impact of CS support programs on participants' build-up of social capital in CS, and the resulting social capital's influence on their persistence in the field, remain unexplored. Motivated by the literature that associates demographic identifiers with access to social capital, and students' access to developmental relationships and career-related resources (social capital) in CS with their persistence, this study explores a CS support program's impact on persistence through capital building. We focus on Google's Computer Science Summer Institute (CSSI), which provided graduating high school students with a 3-week-long introduction to CS. We use interviews with participants who are now 2-5 years out of the program to study CSSI's impact on their social capital and long-term persistence in CS. Thematic analysis reveals three features of the program that influenced students' build-up of social capital, and that the resulting persistence was realized through students' progress towards internships in CS and goals for paying-it-forward in CS. These findings inform our recommendations that future CS support programs and educational settings consider mentorship centered on socioemotional support, opportunities for collaboration, and time for fun social activities. Additional suggestions center on engaging socially-oriented individuals with CS support programs. These insights inform facilitators and educators in CS on design choices that can encourage the persistence of underrepresented students in CS.

## Introduction

The growing digital economy [1] and the widening gap in organizations' access to employees with hard digital skills [2] propel the already rising demand in the Computer Science (CS) profession. According to the U.S. Bureau of Labour and Statistics, Computer Scientists in the U.S. earned a median salary of 136,620 USD in 2022, and the number of CS jobs is expected to grow by 23% in the next decade [3]. With these positive prospects in mind, the number of undergraduate students pursuing CS is rising, too. A study by the National Student Clearinghouse Research Center showed that four-year degrees in *Computer and information sciences and support services* experienced a 34% rise in student enrollment numbers between 2017 and 2022 – more than any other degree [4]. Similarly, the number of Canadians with a degree in *computer and information science* rose by 46% from 2016 to 2021 [5].

While CS careers are lucrative and rising in demand, the profession still suffers from a diversity problem. Data from the U.S. shows women comprising 19% of graduates with bachelor's degrees in CS and less than 30% of graduates with master's and research doctorate degrees [6]. Moreover, the majority of degree holders in CS identify as White, with only 9% and 11% of bachelor degree holders identifying as Black and Hispanic respectively [6]. This breakdown is mirrored in the American workforce: in industry, 65% of CS jobs are held by White employees and women comprise only 28% of the workforce [7].

The need to improve the representation of women and Black, Indigenous, and people of color (BIPOC) in CS motivates studies of young adults' persistence in CS and associated fields. Notably, the understanding that diversity promotes knowledge production in scientific contexts [8] and possesses great relevance in the field of CS, where algorithmic solutions are vulnerable to gender and racial biases [9], [10], reinforces the need for better representation. This also motivates organizations to host education and workforce support programs for youth or young adults from underrepresented backgrounds with an interest in CS. Examples include camps [11], [12], [13], scholarships [14], [15] mentorship projects [16], [17], [18], job-training [19] and college preparation programs [20]. Major corporate organizations have also hosted CS support programs, including Microsoft's Technology Education and Learning Support (TEALS)[1], Google's CS Research Mentorship Program (CSRMP)[2] and Meta University[3], among others.

Motivated by prior works' calls for additional research on effective diversity programs in technology [21] and the links between programs' design choices and students' affective outcomes [22], our work investigates how specific features of a CS-specific support program contributed to the social capital and persistence in CS of students whose identities are underrepresented in CS. More specifically, we investigate the impact of students' participation in Google's Computer Science Summer Institute (CSSI): a 3-week-long program where graduating high school students from historically underrepresented groups in CS acquire technical skills, learn about the field, and grow their network of peers and mentors. In fact, the focus of this work is the last goal: the extent to which participants grow their network in CS. Our attention on social capital is further encouraged by the impact of demographic identifiers on individuals' access to social capital [23], [24], and the recent work on support programs that demonstrate social capital's strong association with alumni's progress towards career goals [25].

To study the social capital that students acquired from CSSI and the relationships with their persistence in CS, we use semi-structured interviews (n=16) with past participants of the program. These interviews reveal qualitative insights into the features of CSSI that students perceive to be most responsible for their build-up of social capital, as well as how the resulting social capital influenced school and/or career decisions in CS. With an understanding of CSSI's impact on students' persistence in CS, our findings will inform the design of future CS support programs such that they encourage and build social capital and persistence among underrepresented students in the field, ultimately working towards greater diversity in CS.

**Background**

*Persistence in CS*

Persistence refers to an individual's commitment (whether intended or realized) to stay in a field. Its study is motivated by the substantial opportunity cost that results when undergraduate students leave the field [26] and CS' highest attrition rates among Science, Technology,

---

[1] https://www.microsoft.com/en-us/teals
[2] https://research.google/outreach/csrmp/
[3] https://www.metacareers.com/careerprograms/pathways/metauniversity

Education, and Mathematics (STEM) fields [27], despite the rise in its post-secondary enrollment numbers [28], [29]. Those who leave CS programs are more likely to be women than men [30], [31], [32]. Further, the culture of CS and systemic barriers in CS compromise the persistence of Black students [33]. While studies suggest that students' demographics, academic experiences, self-assessment of abilities, as well as institutional factors impact persistence, the research on the persistence of underrepresented students in CS remains limited [32]. We strive to add to the literature on persistence in CS and inform the design of CS support programs such that they facilitate the persistence of underrepresented students in the field.

Intentional persistence measures an individual's perceived commitment to remaining in a field into the future [34]. This includes intentions to complete a CS major [26], [35], [36] and/or pursue a career in CS [26], [36], [37]. Shorter-term intentions include students' perceived likelihood that they take another CS course [31]. Based on the work of Cech et al. [34] for engineering, the present study defines intentional persistence to be an individual's commitment to remaining in the field of CS in five years, whether in an academic or professional capacity.

Then, behavioral persistence is an objective measure of whether an individual stays in a field [34]. Lehman et al. [32] follow the persistence of first- and second-year undergraduate students on an annual basis for five years and consider students who remain in a CS major to be *computing persisters*. Other examples of behavioral persistence include taking another course [31], enrolling in and graduating from a Master's or PhD program [38], and staying in a professional position [39] in CS. In the present study, participants who exhibit behavioral persistence are those who currently study CS and/or hold a professional position in the field.

  *CS Support Programs*

CS support programs include summer camps, afterschool programs, workshops, and seminars that span from one to four weeks [22] and are offered by educational institutions, non-profits, technology companies, among other hosts [40]. They commonly target participants in the middle school age group [40], [41] given its significance for the development of identity and interests [40]. Still, programs that focus on high school students are popular too, to promote their pursuit of CS education [40]. Beyond K-12, there are programs that focus on undergraduate students to boost retention in CS [14], [15], especially among underrepresented students [14]. Nonetheless, not all CS support programs target a narrow age group: one program that targeted students in the broad age range of 10 to 18 years old reported the number of participants with a strong interest in cybersecurity doubling after the program [42]. For our study, however, all subjects had already applied to post-secondary studies by the time they started the CS support program. Likely having gone into the program with an interest in CS, we expect the program to influence students' decisions about internships and/or postgraduate pursuits.

While diverse in their approaches, *broadening participation of underrepresented groups* and *developing participants' interest in CS* are the most popular substantive aims of these support programs [40]. It is possible that *the development of students' knowledge* and *skills in CS* are less popular goals because they are deemed better suited for formal education [40]. Nonetheless, support programs commonly strive to provide students with hands-on programming experience, all while enhancing self-efficacy, exposure to the field, and sources of social support [22].

With regards to these programs' coverage of computing topics, general computer programming is most common, followed by mobile apps, making games, and robotics [40]. Meanwhile, their structure usually involves a period of direct instruction at the start, to be followed by a period of guided or open tasks, and a personal project or presentation to conclude the program [40]. To provide students with support, programs may feature CS experts, mentors, and/or role models [22]. Additional program features include field trips, live sessions, and/or videos that introduce students to CS careers to fuel their interest and combat negative stereotypes [22].

*Social Capital*

Social capital refers to one's access to valuable resources that are sourced from social relationships and useful in advancing life goals once mobilized [25], [43]. The types of relationships from which individuals can source social capital include peers (e.g., classmates) and near-peers (e.g., coaches) from mentorship programs [25], religious and familial affiliations, professionals from volunteering and networking events, and faculty members [44].

Many studies investigate the impact of social capital on education and career outcomes. One study finds that the size of students' information support networks of faculty and staff are significant predictors of GPA and that students with higher GPAs have larger peer information support networks [45]. Similarly, Boat et al. [25] find associations between young adults' peer and near-peer social capital and their progress toward their education and career goals.

Social capital may be subdivided into 1) the strength and quality of one's relationships (developmental relationships) and 2) the extent of resources (e.g., information, skills, industry/academic connections) that are accessible from the relationships [25]. The literature on persistence in CS speaks to the impact of developmental relationships and resources. Studies find *positive interactions with peers* to be the strongest predictor for students' intention to major in CS [35] and *peer support* to predict CS degree attainment [46]. Similarly, studies attribute social capital in the form of a parent in CS [32], strong friendships and parental support [47], and encouragement from family and non-family [48], to decisions to pursue and persist in CS.

The significance of social capital is also apparent in studies that focus on underrepresented ethno-racial identities in CS. For example, social capital, in the form of student-led clubs and organizations, mentor networks, and/or financial support, among other resources, is associated with Black students' persistence in CS [33] and engineering [49]. Another study finds that Latina students utilize their support network of peers and mentors to navigate challenges in CS [50]. Together, these studies justify the significance of social capital for students' persistence in CS.

*Social Capital from CS Support Programs*

CS support programs prioritize and enhance participants' social capital. One such example is a tiered mentorship program for underrepresented students that involves faculty, graduate students, peers, and events to discuss shared experiences with other students [14]. Other programs also provide low-income and underrepresented students with role models and mentors in computing [11], [17], [20], and suggest increases in students' self-efficacy [12] and interest in CS [17], [20] after the mentorship. Outside of mentorship, participants can also leave programs with a stronger professional network. One study demonstrates that young African American men leave a program with an increased number of peers who they view as being technical resources [19].

Nonspecific to CS, one study explores education and workforce support programs that connect young adults with support for job applications, on-campus resources, and/or technical skills, where most participants identify as female gender and Latinx, Asian, or Black race [25]. It finds that individuals' peer and near-peer social capital from the programs are significant predictors of their self-initiated social capital: the extent to which they mobilize their network in the pursuit of career goals [25]. Then, the resulting self-initiated social capital is found to be a significant predictor of individuals' progress towards career and education goals [25]. This reinforces the value of support programs' common focus on networking and relationship-building outcomes.

Together, prior findings that suggest support programs enhance students' social capital and social capital's relevance for persistence in CS motivate our study into whether students acquired social capital from CSSI, how CSSI's design and environment influenced students' access to social capital, and whether participants' outgoing social capital impacted their persistence in CS.

- *RQ1: What form(s) of social capital in CS did students acquire from CSSI?*
- *RQ2: What were features of CSSI that influenced students' access to social capital in CS?*
- *RQ3: How did access to social capital from CSSI impact students' persistence in CS?*

## Methods

Hosted by Google and first launched in 2008 [51], CSSI's name recognition and maturity enabled access to a large pool of candidates. Beyond the recruitment advantages, CSSI was aligned with our focus on underrepresented students' persistence, since they targeted "students from historically underrepresented groups" in CS [52]. Their participants would have entered CSSI just after graduating high school, before starting post-secondary studies. CSSI was three-weeks long and held in-person across multiple American cities in cohorts of thirty students, until it moved to being online in 2020 because of the Covid-19 pandemic. Before 2020, most participants stayed in dorms. Google handled living and transportation arrangements.

### *Recruitment and Demographic Survey*

The first author used LinkedIn for recruitment using keyword searches for 'Google Computer Science Summer Institute', 'CSSI', and 'Google CSSI' in November 2023. Those contacted were directed to an expression of interest survey for a subsequent interview. Of the 39 participants who expressed interest, the first 34 were contacted, and 16 scheduled an interview. Participants also self-reported their race, gender, and first-generation status on a demographic survey (see **Table 1**). The questions for race and gender were taken from a guide on inclusive language [53]. Survey and interview protocols were approved by the university's research ethics board. Note that this study is not associated with Google.

### *Interview Protocol*

All 16 interviews were one-hour long and conducted over Zoom. 15 interviews were audio-recorded. One participant did not consent to being audio-recorded. For their interview, the second author typed live notes while the first author asked the questions. While this resulted in fewer direct quotes from their interview, all of their sentiments were recorded and included in the analysis. The interviews consisted of five sections. First, we asked about participants' undergraduate program(s), when and where they participated in CSSI, and their general thoughts about CSSI and its impact. We also asked about their current role(s) and whether they perceive them as being in the field of CS to gauge participants' behavioral persistence in CS (i.e., whether their persistence was realized, years after CSSI). Of the next three sections, one focussed on students' social capital in CS, while the other two investigated students' confidence in their technical abilities and their opinions on CS careers. The final section asked students whether they intend to stay in CS in five years (see **Table 2**). Since the interviews were semi-structured, follow-up questions were asked and minor modifications were made depending on the emergent discussion. While all five sections were incorporated into the analysis, the investigation was focussed solely on participants' build-up of social capital from CSSI. The relationship between the strength of social capital acquired from CSSI and delivery format (i.e., in-person v. remote cohorts) is outside the scope of this work.

*Participant Characteristics*

Of the 16 participants, 11 took part in CSSI between 2018-2019. The remaining 5 participated in remote cohorts between 2020-2021 (see **Appendix A** for more details on participants' cohorts).

Table 1: Participants' demographics and CSSI cohort

| Factor | Value | # | Factor | Value | # |
|---|---|---|---|---|---|
| **Gender** | Women | 10 | **First-generation college student** | Yes | 4 |
| | Men | 6 | | No | 12 |
| **Race/Ethnicity*** | Asian | 4 | **CSSI Location** | In-person | 11 |
| | Black or African American | 4 | | Remote | 5 |
| | Hispanic or Latino | 6 | | | |
| | White | 3 | | | |

*One participant identifies with two races.

Table 2: Participants' behavioral and intentional persistence in CS.

| Current Role (Indicator of Behavioral Persistence) | Participant IDs | Intended Role in Five Years (Indicator of Intentional Persistence) | Participant IDs |
|---|---|---|---|
| Software engineer (incl. interns) | 2, 5, 6, 7, 8, 9, 11, 12, 13 | Software engineer | 2, 5, 6, 12, 13, 7, 1, 10 |
| Security analyst | 14 | Security analyst | 14* |
| Data analyst | 15 | Data analyst | 15, 14* |
| User experience designer | 3, 4 | User experience designer | 3, 4 |
| Sales engineer | 16 | Sales engineer | 16 |
| Management (incl. product and project management) | | Management (incl. product and project management) | 8, 11, 14* |
| Graduate studies in CS | 1, 4 | Graduate studies in CS | 4, 9 |
| Undergraduate student in CS | 2, 5, 10, 11, 13 | Undergraduate student in CS | |

*Participant #14 was unsure about their intended role in five years. They mentioned three possibilities.

*Thematic Analysis*

The 16 interview transcripts were uploaded into the NVivo 14 software for analysis. The first author completed the analysis, following the Braun and Clarke [54] method. In Phase 1, the first author familiarized themselves with the transcripts, taking notes on potential patterns after each interview was complete, and then again after reviewing all transcripts. Once all transcripts were reviewed, the first author proceeded to generate initial codes that included people and relationships, elements of CSSI's curriculum, environments, as well as individuals' attitudes, characteristics, and decisions. For example, this quote was coded with "discord server", "CSSI peers," "casual conversations", and "bonding": "The CSSI server isn't just professional… Conversing with them in the server in a non-professional way built stronger bonds."

Phase 2 concluded with 101 codes. The structure of the 101 codes was multi-level. For example, the 'job application process' code was parent to lower-level codes, such as 'interview preparation' and 'application updates'. To facilitate the search for themes in Phase 3, the first author organized the 101 codes into themes with consideration of the research questions. This also involved revisiting the transcripts to add, revise, and contextualize code assignments for an informed search for themes. In total, 15 themes were formed: 4, 7, and 4 themes that addressed RQ1, RQ2, and RQ3 respectively. In Phase 4, the initial thematic map was presented to and discussed with the second author, which inspired themes to be re-investigated, combined, and discarded. The interview transcripts were also revisited to inform the reorganization of codes into the updated set of themes. For example, themes whose associated excerpts were weak or unclear were discarded if another review of the transcripts was still insufficient in revealing any meaningful data. The final thematic map consisted of 2, 3, and 2 themes that addressed RQ1, RQ2, and RQ3 respectively. See **Appendix B** for more details on the themes.

**Results**

*RQ1: What form(s) of social capital in CS did students acquire from CSSI?*

*Developmental relationships*: During CSSI, students received encouragement from their instructors. Participants #2-3, #10, #13, describe that instructors were "very encouraging", "motivating", and consistently offering "a lot of praise." Then, once CSSI was complete, students sourced emotional support from each other during professional challenges. For Participant #1, their community of CSSI peers gave them the "feeling [that] you're not in this alone" while they struggled with internship offers and technical interviews. Likewise, Participant #12 said that "[it] was nice to have a little community where you could complain about stuff that wasn't working out." For Participant #9, their peers from CSSI would challenge their insecurities, constantly telling them that they were capable of becoming a software engineer.

Participant #8-9 and #11-12 explain that the support also took the form of praise and/or birthday celebrations. In the event of job success, Participant #9 mentioned that "everyone… [would be] hyping each other up." A subset of participants also referenced an instructor who continues to initiate check-ins, demonstrating that they have "stayed invested in [their students'] success."

*Resources from developmental relationships*: Many peers would share early career and college resources. For the former, Participant #1, #4-8, #12, #14-16, mention the "exchange of job opportunities", invitations to and the formation of teams for hackathons, "a list of recruiters", collective leetcode practice over Discord, mock interviews, and a shared Google sheet of internship postings. As for college resources, help about course selection and introductory CS content were mentioned by Participant #8 and #14-16.

Referrals and Google connections that were sourced from the instructors were also mentioned. In the case of Participant #12, they connected with a Google recruiter at their college over having a mutual friend: one of the instructors from their CSSI cohort. For Participant #13, it was one of their instructors giving them a referral for a Google internship. Finally, students sourced career inspiration and insights from their network of peers and instructors from CSSI. For some, this started during the program, as Participant #4 describes, "I learned a lot from their approach… what their interests [were]... technical… design… front-end development… We had a lot of different interests and we shared them." Having these peers on LinkedIn now, Participant #4 benefits from "[the] visibility to what these kinds of people are doing... [and] what they think [are the] most updated [trends] in the industry."

*RQ2: What were features of CSSI that influenced students' access to social capital in CS?*

*Support and encouragement*: Support from the instructors during CSSI resulted in students acquiring tangible skills and career insights, as well as feeling valued and empowered during the program. For example, one instructor approached Participant #12 during the final project, upon noticing that Participant #12 had left their team. As a result of the instructor's advice to be more assertive in their request to contribute, Participant #12 resolved their team conflict and acquired skills in interpersonal communication:

> [He] took me aside… [and said] "You have to stand up for yourself, and you have to actually contribute and stuff cause you know you can."… So I kind of actually went back to the project, and because [my teammate] wasn't letting the other girl contribute either… We actually sat him down and were like, "Look, this is actually like a problem"... That was definitely also like an introduction [to] solving interpersonal conflicts…

Likewise, Participant #10 appreciated how their mentor approached technical support during the final project. Instead of "solving the problem" for them, the mentor would lead Participant #10 and their team *towards* the right direction. This was "more useful for the learning experience."

Participants, such as Participant #14, described that their instructors would go "above and beyond." For Participant #14, this was evident when they would stay to eat lunch and engage with the students outside of their teaching hours. Similarly, Participant #5 said that the instructors "definitely tried to build a connection with their students" and Participant #15 said they were "very available, very open to helping with both the project… and then also general questions."

*Collaboration:* Our participants' accounts provide insights into CSSI's collaborative elements and suggest that they contributed to students' build-up of sources of support in CS. Participant #7 described that, "[they] were sitting in little pods next to each other [as they listened] to a lesson… [Being] in proximity to each other… [made it] so easy to ask questions… [and] get feedback." For Participant #15, the opportunities for collaboration were consistently met with their peers taking their ideas seriously, strengthening their confidence in their ideas as a result:

> Collaborating with other multiple people… I think it made me feel good about my ideas being heard and like being like worth listening to… Every time I said something, like my group members took it seriously and like, even though I wasn't always like right about like what I was proposing… They would still consider it.

Further, Participant #8 and #16 mentioned that they formed close, long-lasting bonds with their teammates from the final project. As Participant #16 describes, this may be because "[they] would naturally be spending more time with [their teammates] than other people in the group." Both of them participated in CSSI more than four years ago. Despite this, "there [are] 3 or 4 [participants]" who Participant #16 still interacts with "on a semi-regular basis", and Participant #8 visited one of their past teammates who works in a different state, earlier this year.

*Social activities*: Students credited social activities to long-lasting friendships from CSSI. Participant #14 attended CSSI more than four years ago and yet, "a lot of the people who [they] did CSSI with are some of [their] best friends today. [They] influence each other… If a specific job area sounds cool, [they'll] do it." These friends were not their teammates: they bonded during activities out in the city and in the dorms:

Yeah, actually, now that I think about it, none of them were my group members… It happened to be the people who wanted to explore the most and do the most outside of our day at Google… We kind of just like would explore [the city] a lot and go out to eat dinner. … Or just even hang out in the dorm.

Participant #12 also attributed the shared living space to bonds with peers who would later share insights about interviews and job offers. They add that career conversations are "awkward to talk about with… other people… but [they] were all living together… [they] really bonded." For Participant #16, playing basketball at CSSI encouraged bonds. Beyond friendship, the experience eased Participant #16's concerns about how their interest in sports aligns with the CS image:

To know them on a more personal level, just really reinforced the fact that, you know, where they're at in CS is only one facet of who they are…So we would go play [basketball] and obviously not talk about anything computer science related… That, I think, helped create those bonds…because there's that whole stereotype in tech that like being in tech and being athletic… doesn't really have a lot of overlap...I've played a lot of sports growing up…that identity clash, I think, [it] was alleviated a lot by doing that with them.

Participant #6 also referenced "hanging out in the common room", "[a] gym", "[a] soccer field", as well as a "really funny" icebreaker activity. More than five years out of the program, Participant #6 still holds three friendships from CSSI. Meanwhile, Participant #3 credited a 3-hour-long walk with an Instructor and two peers for the forming of a mentorship relationship and explained that, "the beauty was in the unscripted moments." Further, Participants #6, #11, and #15 credited the program's length. As Participant #11 describes, "[they] were practically glue for three weeks."

Notably, participants quoted virtual social elements less frequently. Nonetheless, Participant #5 did credit the active engagement in live chats to their remote cohort being, "pretty connected." Then, Participant #1 mentioned that two, organized Zoom reunions enabled them to, "learn a lot about what other people are doing [now]."

*RQ3: How did access to social capital from CSSI impact students' persistence in CS?*

*Internships in CS:* Multiple participants credit their social capital from CSSI for their first internship: a milestone that most agree is daunting to reach without support given the competition in CS. For Participant #12, CSSI "jump started [their] career" because it informed them of Google's STEP and resulted in their connection to a Google recruiter:

[CSSI] was also like an "in" with the recruiters... The recruiter that came to [my college], she actually knew one of the people that was in [my CSSI cohort]... [My] 2 Google internships [were] definitely influenced by that…I know people that never got an internship…even though they're perfectly capable…

Other participants credit the developmental relationships with their peers from CSSI for their internships in CS. For example, Participant #5 credits the influence from a student-led Discord server that hosted leetcode sessions and promoted job opportunities for their current internship:

I think the main benefit... I think it was the people… We are still in touch, even years after… they would talk about internships, they would do the leetcode sessions on the Discord server… I didn't really know anybody in CS… If it weren't for them, I wouldn't have done leetcode as early as I did. I wouldn't have applied to the internships as early as I did. I probably wouldn't even have the internship that I have now, if it weren't for them… They're obviously very career-driven and, you know, that kind of rubbed off on me.

CSSI peers' positive influence on persistence was also referenced by Participant #15: "I might not have gotten my first internship without interacting with other people who are also highly motivated and like doing leetcode…" Notably, the influence enabled resilience, too. Participant #5 mentioned later that their peers helped them "overcome [their] first rejections" by directing them to useful resources, such as leetcode questions and recruiters' contact information.

*Paying-it-forward in CS:* For Participant #1, an instructor remained a source of inspiration after CSSI. This instructor is "always checking in" on their past students and runs "lots of online classes." This instructor is a role model since their dedication to teaching in CS influenced Participant #1 to become an after-school tutor for eighth graders: one step towards their goal of bringing CS education to underrepresented communities, following their instructor's example:

> [They] were [an] engineer… [and they were] here like teaching a bunch of high school students how to code for the first time. I thought that was a cool thing. And honestly, that's something that I would like to do in the future. One of my really big goals is to be able to bring computer science education to more underrepresented communities just because I never really had the opportunity to learn computer science in middle and high school… So, yeah, that's like a really big influence on me. And that's also why I became a tutor now.

While Participant #1 was from an in-person cohort, Participant #10 – a remote participant – had a similar experience. Participant #10's appreciation for their Instructors' enthusiasm for CS inspired them to share the same energy and advice with others in CS:

> They were always very energetic and enthusiastic about what they did... That just reminded me that [while] things [can] be tough in CS… in the end, it's something that I like doing, and that will make me continue doing it… and if I were able to give that energy to other people, that would be really nice… I try to give back to other people by giving them advice and talking to them about my experiences in a similar way.

## Discussion

In response to our RQs, we note three key findings. First, students from in-person and remote cohorts sourced both dimensions of social capital from CSSI: developmental relationships and resources in CS. Second, social capital from CSSI encouraged students' pursuit of internships and their motivation to pay-it-forward in CS. Third, students credited *support and encouragement* from instructors, opportunities for *collaboration*, and *social activities* during CSSI for their build-up of social capital. We contextualize our findings in past work within the categories of *social capital*, *persistence*, and *features of CSSI that built social capital*.

*Social capital*: Our finding that students sourced social capital in CS from CSSI is aligned with other studies whose subjects leave CS support programs with more mentors and role-models in computing [11], [20], peers that can act as technical resources [19], and knowledge and skills in computing and leadership, with a differential effect for students that received mentorship [55].

*Persistence:* Broadly, our demonstration that students' social capital enabled their persistence in CS aligns with prior work that found students' peer and near-peer social capital from non-CS specific support programs to be (either directly or indirectly) positively correlated with their progress towards education and career goals [25]. Other studies on CS support programs that emphasize community building and mentorship also found improved retention of students in CS [56] and computing identity in their participants [15], [55], [57].

We extend prior work to demonstrate that the longitudinal impact of support programs on persistence may be realized by the pursuit of internships. Specifically, participants described that collective leetcode practice and the exchange of job postings encouraged their success with opportunities. Finding that students' social capital motivated their professional development is especially significant when we consider that students who acquire internships in CS demonstrate active engagement in application and preparation processes [58]. Ultimately, students' success with and exposure to internships likely impacts their long-term persistence, and our findings demonstrate that this may be influenced by their social capital from CS support programs.

With specific regard to paying-it-forward in CS, our results demonstrate CS-specific examples of students being inspired by their mentors and/or instructors to continue in CS with socially impactful goals. Thus, our findings provide a CS context for the work of Boat et al. [25]: a quantitative study on the impact of young adults' peer and near-peer social capital from non-CS specific support programs that found students' near-peer social capital to be directly and positively correlated with their resulting 'commitment to paying-it-forward' (e.g., "I pass on my knowledge and skills to others.") [25, p. 1293]. Likewise, another study suggested that their on-campus mentorship program encouraged students' formation of support groups (e.g., *Women in CS*), engagement with the department, and promotion of CS to other students [56].

*Features of CSSI that built social capital:* First, the theme of *support and encouragement* demonstrated how the instructors' commitment to mentorship led to their students feeling valued and confident in their development of new skills in CS. These findings are echoed in the literature that finds mentorship in CS to provide students with skill-building [55], [59], confidence in their abilities [55], [57], improved computing identity [57], sense of belonging [57], and encouragement [59]. The significance of this reassurance is also mirrored in the work of Davis et al. [16], where they study a support program whose mentors were trained to emphasize socioemotional support: displays of encouragement, empathy, and role modeling towards their mentees. They found that the program had a significant, positive impact on students' build-up of confidence, skills, sense of belonging, and research capital in CS [16]. Beyond reinforcing the significance of socioemotional support, our findings suggest how the support can encourage persistence. Specifically, this occurs with participants being inspired to follow their instructors' example to be sources of support for others in CS.

Next, the theme of *collaboration* was also found to be beneficial for students' formation of bonds in CS. This result is reflected in prior work whose results suggest that the long-term impacts of project-based learning in STEM transcend traditional learning outcomes to also include professional advancement and friendships [60]. Further, authors demonstrate that students' exposure to collaborative assignments are a significant, positive predictor of their persistence in CS [26]. Interestingly, however, the more recent work of Lehman et al. [32] found that students' exposure to collaborative pedagogy in introductory CS courses was a significant, negative predictor for persistence. In their discussion, they suggest that the surprising result may indicate the strong influence of the *quality* of collaboration received, hence their recommendation for future works to qualitatively inquire into how collaboration is organized in CS education [32]. From our findings, it is plausible that certain features of CSSI, such as "so much access to those instructing [them]" and "being able to learn without being pressured for grades" enabled the collaboration to be fruitful for students' involvement in the profession. This reinforces the recommendations by Schulz et al. [61] for successful collaboration in CS education, specifically their mentions of adapted assessments and concrete guidance from teachers.

The last theme, *social activities*, was credited for close bonds with peers who would later become friends and professional allies. Examples were playing sports, rooming together, and exploring CSSI's host city. While CS support programs can incorporate "informal experiences [for] creating a sense of cohort solidarity" [14, p. 15], our findings suggest that these activities can also have a long-lasting and positive impact on students' network in CS. For CSSI participants, this included multi-year-long friendships and professional relationships.

**Implications**

*Inspire social benefit interest in CS:* Our findings suggest that students' social capital from CS support programs may inspire their commitment to paying-it-forward in CS, thereby encouraging their persistence. Given that a desire for social impact has been negatively associated with persistence in CS [48], [62] and AI/ML [63], CS support programs may be a promising opportunity to further engage and motivate socially-oriented students in the field. As suggested by previous work for AI/ML [64], retaining socially-oriented students may also promote gender diversity since women show higher levels of 'social benefit interest' than their peers who do not identify as women [63].

While our study demonstrated that inspiration from instructors' displays of support and enthusiasm for CS inspired students' plan for social impact in CS, we also acknowledge that CS support programs have the potential to better address socially-oriented interests. This may be achieved by introducing students to speakers in CS backgrounds with obvious social impact, such as healthcare [64], and/or engaging students in service learning opportunities that result in the "sharpening [of their] … skills and CS identity." [14, p. 14].

*Encourage student-led communities:* In response to, "How can we prepare students [that lack the agentic resources to secure internships in CS]… so that they have the necessary skills to thrive in the job recruitment process? [58, p. 6], we found that students' social capital from CS support programs may be mobilized to advance their success with internships in CS. We believe that future CS support programs should encourage their students to stay connected once the program is complete for access to a network of professional information and motivation in CS.

*Add fun to the agenda:* Future programs should incorporate opportunities for fun to encourage students to socialize and form bonds. Sports may be especially beneficial, given that they "can be a generative space for engaging youth in meaningful learning with and about tech." [65] From one participant, we understand that playing sports at CSSI alleviated their negative stereotypes about CS and encouraged close bonds with their peers by humanizing them beyond their CS skills. Nonetheless, fun can be achieved in virtual settings too. One study adapts the games of an in-person CS support program to succeed in "keeping the fun alive" virtually [66].

**Limitations and Future Work**

Due to the small sample size, we cannot assume our findings to reflect the experiences of all participants. Further, the recruitment method was subject to bias. Past participants of CSSI that found the experience to be ineffectual may have been less likely to add CSSI to their LinkedIn or opt into our study. It follows that this work does not position itself as an evaluation of CSSI's effectiveness. Instead, we use our participants' accounts to reveal how CSSI influenced at least a subset of students' social capital and persistence in CS, and suggest advantageous design elements for future programs. We also note that participants' accounts of CSSI may have been subject to recall bias, especially given the years since their participation. Nonetheless, the time delay enabled insights into the long-term impact of social capital and its influence on persistence.

We also acknowledge that the coding for the thematic analysis was conducted only by the first author. While this ensured consistency in code assignments, they would have been vulnerable to the first author's biases. We note, however, that discussions with the second author did motivate additional iterations and revisions to the coding scheme. Additionally, we did find support for our findings where the significance and implementation were discussed in past literature on CS education. Our contributions contextualized other researchers' quantitative findings with CS-specific examples and/or demonstrated their relevance to social capital and persistence in CS.

In future work, we will extend the work of Boat et al. [25] on education and workforce support programs with a quantitative study that investigates changes in participants' social capital, and relationships with persistence in CS. The study will incorporate a survey to validate the findings of this work, and control for variables that were outside its scope, including the time elapsed since participants' completion of the CS program [67] and location of participation [25]. We will specifically investigate the impact of remote v. in-person delivery. While our results showed that all subjects, including the five remote participants, acquired social capital from CSSI, research suggests that students perceive face-to-face classrooms to be conducive for stronger bonds with their peers [68]. However, another study reported on a CS support program that converted to a remote format while maintaining students' satisfaction levels [66].

For a holistic understanding of CS support programs' impact on affective outcomes, we will also include measures that have been previously associated with persistence in STEM, such as technical confidence [31], [32], [36], [69], [70] and professional role confidence [34], [63], [71]. Further, we will be interested in how social capital influences the relationship between students' perceived performance/competence in CS and persistence. Performance/competence is theorized to be an advanced measure of self-efficacy [36] – also linked to student retention in CS [72] – and shown to have a direct effect on students' interest and persistence in CS [36]. Finally, we will also add social-benefit interest, motivated by the findings of this study and research on persistence in AI/ML [63].

**Conclusion**

We studied a CS support program for incoming college students from underrepresented backgrounds to understand how its features influenced social capital and persistence in CS. From sixteen interviews with past participants of CSSI from the years 2018-2021, we demonstrated the significance of instructors' socioemotional support, collaborative activities, and non-organized opportunities for fun. Backed by the literature, these findings inform CS educators and program coordinators on features that encourage students' build-up of skills and bonds in CS.

For impact on persistence, students' social capital enabled their successful pursuit of internships, and motivated their intentions for social impact in CS. We use these findings to suggest that CS support programs encourage their participants to form student-led, virtual communities for professional development, and engage socially-oriented individuals, respectively. Above all, we encourage facilitators of CS support programs and educational settings to remain strategic in the design of curriculums and environments. With an emphasis on changes in systems over students [26], [73], we believe that the field can resourcefully nurture the persistence of students of diverse identities, ultimately encouraging a future of equitable innovation in CS.

# Appendix

## Appendix A: The representation of CSSI cohorts among the 16 interview participants.

Table A1: Representation of CSSI cohorts by location and year.

| Factor | Value | # | Factor | Value | # |
|---|---|---|---|---|---|
| **Location** | | | **Year** | | |
| | Mountain View | 2 | | 2018 | 2 |
| | Pittsburgh | 7 | | 2019 | 9 |
| | Seattle | 2 | | 2020 | 3 |
| | Remote* | 5 | | 2021 | 2 |

* CSSI moved from in-person to remote in 2020 because of the Covid-19 pandemic.

## Appendix B: The final thematic map.

Table A2: The themes, codes, and examples for the final thematic map.

| Theme | Codes | Example |
|---|---|---|
| *RQ1: What form(s) of social capital in CS did students acquire from CSSI?* | | |
| Developmental relationships | Inspiration (incl. dedicated, inspiring, success, unique interests, feedback), Emotional support (incl. check-ins, celebrating each other, encouraging, friends, sharing hobbies) | "I feel like if I were to like post on Linkedin about like… moving to Washington and interested in like hearing from people who live there…or like considering changing career fields…I think it's just kind of the general benefit of having a wide network. It's more people that I know [and] people that know me. [It's] people who are like willing to help me and also [people that] I can help." – Participant #15 |
| Resources from the developmental relationships | Early career resources (incl. hackathons, application updates, interview preparation, Google's STEP preparation, google sheet of internships), college resources (incl. course selection, cs content help, essays, assignments, easier transition) | "They gave me help... There was one time where I needed assignment help… and one of them…he helped me with a lot of [my] Java code… That network really helped me get through college in the beginning years of college." – Participant #8 |
| *RQ2: What were features of CSSI that influenced students' access to social capital in CS?* | | |
| Support and encouragement | Instructors (incl. approachable, diverse, easy access, going above and beyond), peers (incl. Academically-driven, career-driven, shared goals), relatable, diversity, welcoming | "[They were] from Mexico, so I was able to relate to [them] in that aspect… and practice my Spanish as well…I felt very well-connected to [them]... [They] helped me a lot during my group project…I might have to reach out to [them] after this [interview]...[They were] a great mentor." – Participant #11 |
| Collaboration | Project, teammates, interactive chat, paired activities, pods | "I think the biggest thing was that there were people that I was in like the projects with because you would naturally be spending more time with those people than other people in the group…" – Participant #16 |
| Social activities | Dorming, non-organized fun (incl. exploring city, pool table, sports), program length, | "It [took] a couple of days [to form our friend group]… There were extra-curricular activities that we would do… There was a gym… a soccer field as well… a pool |

| | yearbook activity, zoom reunion | table….." – Participant #6 |
|---|---|---|
| *RQ3: How did access to social capital from CSSI impact students' persistence in CS?* | | |
| Internships in CS | First internship, Google's STEP | "I didn't get much interviews or offers…I would tell them [and] they would point me to resources… one of them… gave me like a whole list of recruiters… [That] helped me overcome the first rejections when I was first applying to internships to then attaining an internship offer." – Participant #5 |
| Paying-it-forward in CS | Cs tutor, motivated to stay, give back | "I really was drawn to the field [of CS] because I enjoyed getting to collaborate with people to solve problems that just like positively impacted someone and I got that first taste [at] Google CSSI… I got to talk to so many people from different backgrounds from across the whole country… Just [seeing] their passion and [hearing] their why… I think that really helped me see that this was a possibility." – Participant #7 |